\documentclass[aps,prl,superscriptaddress,amsmath,amssymb,preprintnumbers,showpacs,twocolumn]{revtex4}
\usepackage{graphicx,color}
\usepackage{amssymb}
\usepackage{amsmath}
\usepackage{bm}
\newcommand{\tr}{\mathop{\mathrm{tr}}\nolimits}

\newcommand{\HA}{\mathop{\mathcal{H}}\nolimits}

\newcommand{\I}{\mathop{\mathbb{I}}\nolimits}

\newcommand{\bra}[1]{\langle #1 |}
\newcommand{\ket}[1]{| #1 \rangle}

\definecolor{dgreen}{rgb}{0,0.5,0}

\definecolor{delete}{cmyk}{0.5,0,0,0}
\definecolor{deletey}{cmyk}{0,0.5,0,0}

\begin{document}


%
\title{
Determining eigenvalues of a density matrix with minimal information in a single experimental setting}

\author{Tohru Tanaka}
\affiliation{Department of Physics, Waseda University, Tokyo 169-8555, Japan}

\author{Yukihiro Ota}
\affiliation{CEMS, RIKEN, Saitama 351-0198, Japan}

\author{Mitsunori  Kanazawa}
\affiliation{Department of Physics, Waseda University, Tokyo 169-8555, Japan}

\author{Gen Kimura}
\affiliation{College of Systems Engineering and Science, Shibaura Institute of Technology, Tokyo 108-8548, Japan}

\author{Hiromichi Nakazato}
\affiliation{Department of Physics, Waseda University, Tokyo 169-8555, Japan}

\author{Franco Nori}
\affiliation{CEMS, RIKEN, Saitama 351-0198, Japan}
\affiliation{
Physics Department, University of Michigan, 
Ann Arbor, Michigan 48109-1040, USA}


\begin{abstract}
Eigenvalues of a density matrix characterize well the quantum state's
 properties, such as coherence and entanglement.  
We propose a simple method to determine all the eigenvalues of
an unknown density matrix of a finite-dimensional system in a single experimental setting.
Without fully reconstructing a quantum state, eigenvalues are determined
with the minimal number of parameters obtained by a measurement of a
single observable.  
Moreover, its implementation is illustrated in linear optical and superconducting systems. 
\end{abstract}


\maketitle

The eigenvalues of a density matrix are fundamental quantities in quantum
physics, and characterize many quantum properties, such as coherence and
entanglement. 
Typically, a function of the density-matrix's eigenvalues allows us to
examine features of a quantum state. 
The von Neumann entropy, for example, is defined as the Shannon entropy of
eigenvalues of a density matrix, and have different applications such
as thermodynamic entropy\,\cite{ref:vN}, optimal compression rate
of a quantum state\,\cite{ref:OptimalRate}, and entanglement
measure\,\cite{ref:BDSW}. 
Other entropies of eigenvalues, e.g., the R$\acute{\rm e}$nyi
entropy and the Tsallis entropy~\cite{ref:abe} can have curious
applications in thermodynamics and statistical mechanics. 
The entanglement spectrum\,\cite{Li;Haldane:2008} depends on spectrum of a
reduced density matrix, and is useful for studying the ground-state
properties of many-body quantum systems in low dimensions. 
Thus, the eigenvalues of a density matrix (and their functions) are
probes into quantumness in various issues. 

Developing a method for determining eigenvalues of an unknown quantum
state is highly desirable, because one can experimentally test many
theoretical ideas about sensing quantum features.  
A fundamental question here is whether the method is simple enough, and
implemented by a small number of experimental setting,
hopefully by a single setup.
We first consider what is a simple approach. 
Reconstructing a density matrix via quantum-state
tomography~\cite{James;White:2001} leads to determining all the
eigenvalues. 
In a $d$-level quantum system, however, we have \mbox{$(d^2-1)$} numbers of
free parameters to be fixed in the reconstruction, so it is highly redundant to
determine $d$ numbers of the eigenvalues, where only \mbox{$(d-1)$} numbers of free
parameters (with the normalization) are necessary.  
Actually, without a full reconstruction of a quantum state, the
eigenvalues of a density matrix are attainable. 
We only need to know the moments of a density matrix $\rho$, as seen in,
e.g., Ref.~\cite{Kimura:2003}. 
There are experimental proposals~\cite{ref:Brun,ref:Flip,ref:Ekert1} to
directly measure $\tr \rho^{k}$ ($k=2,\ldots,d$). 
Since the number of the unknown eigenvalues is equal to that of
the measured quantities, this approach is regarded as an eigenvalue
determination with minimum information.  
We call such a method to be {\it minimal}. 
An eigenvalue determination with minimum information is considered to be
simple enough, since there is no redundancy. 
The minimality of information gain is important in
quantum systems, from the viewpoint of the information-disturbance
relation\,\cite{Fuchs;Peres:1996,Banaszek:2001}. 
Extracting minimal information can lead to suppressing unnecessary
disturbance in a quantum state. 

Next, let us consider the implementability of eigenvalue determination,
with a single setup. 
In proposed methods~\cite{ref:Brun,ref:Flip,ref:Ekert1} with the minimality, the
$k$th moment of $\rho$ is measured by the expectation value of an
observable under an identically-and-independently-distributed (i.i.d.) state, 
$\rho^{\otimes k}$.  
Thus, this approach requires $(d-1)$ kinds of the experimental settings to
determine all the eigenvalues. 
A single-setup determination is built straightforwardly by
a measurement of an information
complete positive operator-valued measure (see, e.g.,
Ref.~\cite{DAriano;Sacchi:2002}). 
We recall that the information completeness is defined by the
state-reconstructing ability from statistical measurement data. 
One can reconstruct a density matrix, via a single experimental setting
for such a measurement~\cite{footnote1}. 
However, as pointed out above, this method does not have minimality. 
An alternative way for obtaining all the moments of $\rho$ with a single setup is to use random unitary operations on a single system\, \cite{ref:random}.
This method is applicable to photon qudits passing through a disordered
medium\,\cite{random_2}, although one must guarantee the uniformity of
the random operations. 
\begin{figure}[tb]
\centering
\scalebox{0.45}[0.45]{\includegraphics{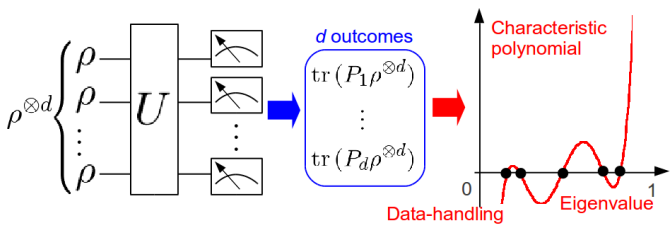}}
\caption{(Color online) Schematic diagram of a single-observable-based method for an
 eigenvalue determination of a density matrix. 
First, one prepares an identically-and-independently-distributed state,
 with finite number copies of a target density matrix $\rho$. 
The number of the copies is equal to the dimension $d$ of a system. 
Next, $d$-outcome measurements with orthogonal projectors
 $(P_{i})_{i=1}^{d}$ are performed by a global unitary
 gate $U$ and local measurements. 
Then, one numerically finds the root of the characteristic polynomial reconstructed
 from the measured data. }
\label{measure_eigen}
\end{figure}

In this article, we propose a method for measuring the eigenvalues of a
density matrix, equipped with both minimality and
single-setup implementability.
Our proposal for a $d$-level quantum system consists of a measurement of
a single $d$-valued observable, or equivalently, a measurement of a
projection-valued measure (PVM) measurement $(P_i)_{i=1}^d$, on a $d$-i.i.d.\ state, $\rho^{\otimes d}$. 
Our central idea is to build a specific unitary gate in a multi-partite system.
We will show that the characteristic polynomial of any density matrix $\rho$ is constructed in terms of probability distributions $\tr(P_{i}\rho^{\otimes d})$ as
\begin{equation}\label{eq:MainResult}
 \det (x - \rho) = \sum_{i=1}^{d} m_{i}(x) \tr(P_{i}\rho^{\otimes d})
\quad
(x\in \mathbb{R}). 
\end{equation}
Here $m_{i}(x) \ (i = 1 \cdots d)$ is a known polynomial of $x$, 
determined according to the choice of the single observable.    
From the PVM measurement $(P_i)_{i=1}^d$ on state
$\rho^{\otimes d}$, we obtain the probability distribution 
$\tr(P_{i}\rho^{\otimes d})$.  
Then, we can calculate all the eigenvalues of $\rho$, via formula
\eqref{eq:MainResult}.  
This process can be efficiently performed by a classical computer, since
the problem is a 1D root finding, simpler than a full
matrix diagonalization.  
Figure \ref{measure_eigen} is the summary of our proposal. 
Moreover, we will show methods for implementing our proposal in physical
systems. 

One of the simple realizations of our approach is to use the
anti-symmetrizer (projector on a fermionic ground state). 
To see this, let us start with some mathematical ingredients. 
The unitary operator on $\HA^{\otimes k}$ associated with a permutation
$\sigma \in \mathfrak{S}_{k}$ over $k$ integers $\{1,\ldots,k\}$ is 
\begin{equation}
U_{\sigma}\ket{\phi_1} \ldots \ket{\phi_k} 
=
\ket{\phi_{\sigma(1)}} \ldots \ket{\phi_{\sigma(k)}},
\end{equation}
for $\ket{\phi_{i}} \in \mathcal{H}$ \mbox{($i=1,\ldots,k$)}. 
An $m$-cycle $c_{m} \in \mathfrak{S}_{k}$ is a permutation to cycle $m$
distinct integers from $\{1,\ldots,k\}$, with others being fixed.  
Noting that $\tr \rho =1$, we have 
\begin{equation}
 \tr (U_{c_{m}} \rho^{\otimes k}) = \tr \rho^m 
\quad
 (m=2,\ldots,k), 
\label{eq:moments}
\end{equation}
for any $m$-cycle $c_{m}$. Thus, each moment of $\rho$ is related to a
physical process.   
An observable for determining $\tr \rho^{m}$ can be constructed by taking 
the hermitian part of $U_{c_{m}}$, as seen in, e.g., Ref.~\cite{ref:Brun}. 
For a general permutation $\sigma$, we can use the unique decomposition
by cycles (see, e.g., Ref.~\cite{Ludwig;Falter:1996}) to obtain 
\begin{equation}
 \tr (U_{\sigma} \rho^{\otimes k}) 
= \prod_{m=1}^k \tr (\rho^m)^{j_m(\sigma)} \,,
\label{eq:GeneralMoments}
\end{equation}
where $j_m (\sigma)$ is the number of the $m$-cycles in $\sigma$. 
For instance, we have 
$ \tr (U_{\sigma} \rho^{\otimes 7}) = (\tr \rho^2)^2 (\tr \rho^3)$ 
for $\sigma = (12)(34)(567)$, because 
$j_2(\sigma) = 2$ and $j_3(\sigma) = 1$, with others being zero. 

The anti-symmetrizer on $\mathcal{H}^{\otimes k}$ is defined by  
\begin{equation}
 A_k
=
\frac{1}{k!} \sum_{\sigma \in \mathfrak{S}_{k}}
\mbox{sgn}(\sigma) U_{\sigma}
\quad
(k=1,\,\ldots,\,d),
\label{eq:gen_proj}
\end{equation}
where $\mbox{sgn}(\sigma) = \pm 1$ is the sign of $\sigma$. 
$A_k$ is a projection operator on 
$\mathcal{H}^{\otimes k}$ ($A_k=A_k^2 = A_k^\dagger$), and has a natural
extension on $\HA^{\otimes d}$ by $A_k \otimes \I^{d-k}$. 
Hereafter, we use the same symbol $A_k$ on $\HA^{\otimes d}$ 
and define $A_0$ to be the identity operator on $\HA^{\otimes d}$. 
Using Eqs.~(\ref{eq:GeneralMoments}) and (\ref{eq:gen_proj}), we obtain 
\begin{equation}
\tr (A_k \rho^{\otimes k}) 
= \frac{1}{k !} \sum_{\sigma \in
 \mathfrak{S}_{k} } \prod_{m=1}^k (\mu_{m})^{j_{m}(\sigma)},
\label{eq:cycle_index}
\end{equation}
with $\mu_{m} = (-1)^{m-1} \tr \rho^{m}$. 
Here we have used 
$\mbox{sgn}(\sigma) = \prod_{m=1}^k [ {\rm sgn}(c_{m}) ]^{j_m(\sigma)}$ and
${\rm sgn}(c_{m}) = (-1)^{m-1}$ for an $m$-cycle $c_{m}$.  
Thus, a projective measurement about $A_{k}$ includes the moments of
$\rho$, up to the $k$th order. 

Now, we show a way to reconstruct the characteristic polynomial of
$\rho$, with Eq.~(\ref{eq:cycle_index}). 
Let us write $\tr (A_k \rho^{\otimes k})$ as $a_k$, and formally define $a_0 = 1$. 
A straightforward calculation of the right-hand-side of
Eq.~(\ref{eq:cycle_index}) leads to the Newton-Girard
formula\,\cite{Dickson:1914} 
\begin{equation}
a_{k} = \frac{1}{k}\sum_{m=1}^{k} \mu_{m} a_{k-m}
\quad
(k=1,\ldots,d). 
\label{eq:reccurence}
\end{equation}
Thus, the sequence of $\{ a_k\}_{k=1}^d$ is equivalent to that of the coefficients of the characteristic polynomial (i.e., elementary
symmetric polynomials).  
To sum up, we obtain 
\begin{equation}
 \det (x - \rho) = \tr [M(x)\rho^{\otimes d}],
\label{eq:char_poly}
\end{equation}
with 
\(
M(x) 
=  \sum_{k=0}^{d} (-1)^k x^{d-k} A_k 
\).  
This result is notable, because the characteristic polynomial is
described by a single quantum observable $M(x)$. 
However, we still need to remove the dependence on the continuous
variable $x$, to make a single-seup approach possible.   
The key is the following structure of the anti-symmetrizers. 
Since a permutation procedure in $A_{k}$ is a part of $A_{l}$ when 
$k<l$ ($\mathfrak{S}_{k} \subset \mathfrak{S}_{l}$), we find that 
\begin{equation}
 A_{k} A_{l} = A_{l} A_{k} = A_{l}
\quad
(k<l). 
\label{eq:relation_proj}
\end{equation}
Furthermore, the projective property of $A_{k}$ leads to an
eigen-subspace $\mathcal{A}_{k} = A_{k}\, \mathcal{H}^{\otimes d}$. 
Therefore, we obtain the inclusion relation for $\mathcal{A}_{k}$: 
\begin{equation}
 \mathcal{A}_{d} \subset 
\mathcal{A}_{d-1} \subset 
\dots \subset 
\mathcal{A}_{2}. 
\label{eq:gen_inclusion}
\end{equation}
We note that $\mathcal{A}_{1} = \mathcal{H}^{\otimes d}$. 
Thus, we have an orthogonal decomposition of $\HA^{\otimes d}$
as $\oplus_{i=1}^d {\cal B}_i$, 
with 
${\cal B}_1= {\cal A}_d $ and
\mbox{
${\cal B}_i = {\cal A}_{d-i+1} - {\cal A}_{d-i+2}$
} 
($i = 2,\ldots,d$).  
The projection operators $P_i$ onto ${\cal B}_i$ are then defined by 
$P_1 = A_d$ and \mbox{$P_i = A_{d-i+1} - A_{d-i+2}$} 
($i = 2,\ldots,d$).  
They constitute our PVM measurements $(P_i)_{i=1}^d$. 
By definition, we find that $A_i = \sum_{j=1}^{d-i+1} P_j$.
Substituting this formula into Eq.~(\ref{eq:char_poly}), we obtain the
practical formula \eqref{eq:MainResult} for eigenvalue determination,
with 
\begin{equation}
m_{i}(x) = \sum_{k=0}^{d-i+1} (-1)^{k} x^{d-k}. 
\end{equation}
Since the dimension of $\mathcal{A}_{k}$ is 
\mbox{$D(k) = d^{d-k}\, _{d}C_{k}$}, 
the dimension of ${\cal B}_i $ (i.e., ${\rm rank}\,P_{i}$) is
\mbox{$D(d-i+1) - D(d-i+2)$}, where we define $D(d+1)$ to be zero.  

We stress that in our approach the use of the anti-symmetrizer $A_{k}$
is not essential. 
For instance, the symmetrizer on $\mathcal{H}^{\otimes k}$ (projector on
a bosonic ground state) leads to a similar formula to
Eq.~(\ref{eq:reccurence}), with the parallel argument above. 
Hence, the symmetrizer works for the eigenvalue determination. 
The virtue of the anti-symmetrizer is a direct connection to the coefficients of the characteristic polynomial with the expectation values of the anti-symmetrizers, as shown in Eq.~(\ref{eq:char_poly}). 

We also notice that the number of target-state copies may decrease if only partial information on the eigenvalues is needed. 
In particular, if the number of (approximately) zero eigenvalues, $d_{0}(<d)$ is known a priori (alternatively can be estimated), 
the present method with $\rho^{\otimes (d- d_{0} )}$ leads to $( d- d_{0})$ predominant eigenvalues. 
Any prior information other than $d_{0}$ is not required.

\begin{figure}[tb]
\centering
\scalebox{0.38}[0.38]{\includegraphics{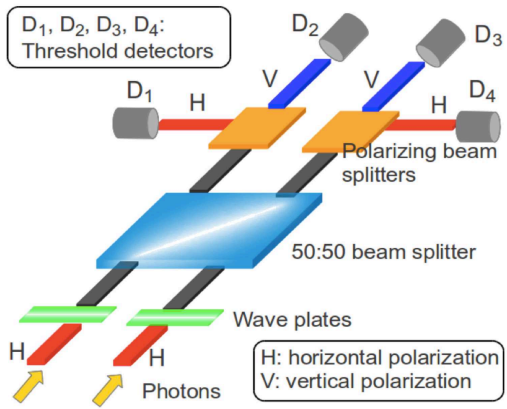}}
\caption{(Color online) Proposal for an eigenvalue
 determination with linear optical qubits. }
\label{fig:bell_analyze}
\end{figure}
Now, we illustrate the present method in physical systems. 
We first consider qubits 
($d=2$ and $\HA={\rm span}\{\ket{0},\,\ket{1}\}$). 
The relevant projectors are  
\(
P_{1} = \ket{\Psi_{-}}\bra{\Psi_{-}} 
\) 
and 
\(
P_{2} = \I^{\otimes 2} - \ket{\Psi_{-}}\bra{\Psi_{-}}
\), 
with 
\mbox{
\(
\ket{\Psi_{-}} = (\ket{01}-\ket{10})/\sqrt{2}
\)}. 
The first example is a linear optical qubit. 
A quantum state is encoded by horizontal (${\rm H}$) and 
vertical (${\rm V}$) polarization (i.e., $\ket{\rm H}=\ket{0}$ and
$\ket{\rm V}=\ket{1}$).  
A two-photon interferometer, the
Bell-state analyzer (see, e.g.,
Refs.~\cite{Braustein;Mann:1995,Michler;Zeillinger:1996}), leads to an
implementation of the projectors, as seen in Fig.~\ref{fig:bell_analyze}. 
Although the main part of the eigenvalue determination is the $50:50$
beam splitter and the subsequent single-photon detectors, this setting
includes a method to prepare a $2$-i.i.d.\ state as well. 
Let us first explain this preparation stage. 
The initial photons are horizontally polarized. 
A series of wave plates is set on each input arm in the interferometer. 
On every trial, one randomly changes a
combination of the wave plates to make an element of 
${\rm SU}(2)$\,\cite{Simon;Mukunda:1990}.  
This procedure probabilistically produces different unitary gates on a polarization state. 
Thus, one can examine the eigenvalue determination of 
\mbox{
\(
\rho = \sum p_{r}V_{r}\ket{{\rm H}}\bra{{\rm H}}V_{r}^{\dagger}
\)}, with $\sum p_{r}=1$, $p_{r}\ge 0$, and $V_{r}\in {\rm SU}(2)$. 
Next, we turn to the determination part. 
After the interference at the $50:50$ beam splitter, either
anti-bunching or bunching occurs~\cite{Braustein;Mann:1995}. 
From the bosonic character of photons, the anti-bunching is related
to an anti-symmetric property with respect to the polarization (i.e., 
\mbox{
\(
(\ket{{\rm HV}} - \ket{{\rm VH}})/\sqrt{2}
\)}), whereas bunching is relevant to symmetric states. 
Hence, using two polarizing beam splitters and four threshold detectors 
(${\rm D}_{1}$, ${\rm D}_{2}$, ${\rm D}_{3}$, and ${\rm D}_{4}$), we
obtain ${\rm tr}(P_{i}\rho^{\otimes 2})$.  
When either $({\rm D}_{1},\,{\rm D}_{3})$ or 
$({\rm D}_{2},\,{\rm D}_{4})$ detects photons, this event is counted as
$P_{1}=1$. 
The others correspond to the case $P_{2}=1$. 

The main idea in our proposal is to use a projective measurement on a
multi-copy of a targe quantum state. 
This technique is applied in photon-qubit
experiments~\cite{Bovino;Sergienko:2005,Walborn;Buchleitner:2006,experiment_3},
to measure {\it non-linear} observabales such as concurrence. 
In other words, the present work studies such
experimental techniques, from the viewpoint of minimality and
single-setup implementability.

Next, we examine a solid-state system, especially a superconducting
qubit~\cite{You;Nori:2005}. 
Let us consider a controlled gate
\mbox{$U_{\rm ZZ} = \exp[-i(\pi/4J)H_{\rm ZZ}]$}, with 
\mbox{$H_{\rm ZZ} = J\sigma_{z}\otimes \sigma_{z}$}. 
The $2\times 2$ Pauli matrices are $\sigma_{x,y,z}$. 
The qubit-coupling Hamiltonian $H_{\rm ZZ}$ is realized in various
systems such as flux qubits\,\cite{Plantenberg;Mooij:2007} and transmon
qubits\,\cite{DiCarlo;Schoelkopf:2009,Poletto;Steffen:2012}. 
This control gate with single-qubit gates leads to
\begin{equation}
U_{\rm D}
=
(\I \otimes U_{\rm H})
(Z_{-\pi/4}\otimes Z_{-\pi/4})
U_{\rm ZZ} (\I \otimes U_{\rm H}), 
\end{equation}
where
\mbox{
\(
U_{\rm H}
=
Y_{-\pi/8}Z_{\pi/2}Y_{\pi/8}
\)}, 
\mbox{
\(
Z_{\theta} = \exp(i\theta \sigma_{z})
\)}, 
and 
\mbox{
\(
Y_{\theta} = \exp(i\theta \sigma_{y})
\)}. 
We find that 
\(
U_{\rm D}\ket{\Phi_{-}} = \ket{00}
\), 
\(
U_{\rm D}\ket{\Psi_{+}} = \ket{01}
\), 
\(
U_{\rm D}\ket{\Phi_{-}} = \ket{10}
\), 
and 
\(
U_{\rm D}\ket{\Psi_{-}} = \ket{11}
\), up to overall phases, where 
\(
\ket{\Phi_{\pm}} = (\ket{00}\pm\ket{11})/\sqrt{2}
\) and 
\(
\ket{\Psi_{+}} = (\ket{01} + \ket{10})/\sqrt{2}
\). 
If both of the qubits are detected as $\ket{1}$ after performing $U_{\rm
D}$, the projector $P_{1}$ is done. 
We refer to the extendability of our proposal useful for the
implementation in general systems. 
Using a superoperator $\Xi$ such that 
\(
\Xi(\rho^{\otimes d}) = \rho^{\otimes d}
\),
and 
its adjoint $\Xi^{\ast}$, 
\(
\tr[\Xi^{\ast}(A)\, B] = \tr [A\, \Xi(B)]
\), 
we find that an expansion of the characteristic
polynomial is not unique. 
In contrast to $M(x)$ in Eq.~(\ref{eq:char_poly}), the observable
$\Xi^{\ast}(M)(x)$ can involve operators other than the
anti-symmetrizers. 
Thus, one may perform our proposal, not sticking to the PVM measurement. 
Let us apply this technique to the eigenvalue determination in a linear
optical qutrit ($d=3$). 
Our qutrit is a superposition of 3-path (or mode) single-photon
states. 
The corresponding bosonic creation operators are $a^{\dagger}_{\ell}$
($\ell=1,2,3$). 
To represent a 3-i.i.d. state $\rho^{\otimes 3}$, we need two additional spatial modes
$b^{\dagger}_{\ell}$ and $c^{\dagger}_{\ell}$, each of which has another
spatial-mode-index $\ell$ for expressing a qutrit state, like
$a^{\dagger}_{\ell}$. 
To simplify the notations, we will denote $b_{\ell}^{\dagger}$ $(c_{\ell}^{\dagger})$ as $a^{\dagger}_{\ell +3}$ ($a^{\dagger}_{\ell +6}$).
In our setting, a 3-i.i.d. qutrit enters an interferometer.
For each $\ell$, a mixing among $a_{\ell}^{\dagger}$, $a_{\ell +3}^{\dagger}$, and $a_{\ell +6}^{\dagger}$ occurs,
\begin{equation}\label{qutrit_1}
\begin{pmatrix}
a_\ell^{\dagger} \\ 
a_{\ell+3}^{\dagger} \\ 
a_{\ell+6}^{\dagger} 
\end{pmatrix} 
\to 
\frac{1}{\sqrt{3}}
\begin{pmatrix} 
1 & 1 & 1 \\ 
1 & e^{i\frac{2 \pi}{3}} & e^{i\frac{4 \pi}{3}} \\
1 & e^{i\frac{4 \pi}{3}} & e^{i\frac{2 \pi}{3}}
\end{pmatrix}
\begin{pmatrix}
a_\ell^{\dagger} \\ 
a_{\ell+3}^{\dagger} \\ 
a_{\ell+6}^{\dagger} 
\end{pmatrix} .
\end{equation}
This transformation (a qutrit quantum Fourier
transformation~\cite{Tabia:2012}) does not alter the bosonic canonical
commutation realtions, and can be built up by beam splitters and
phase shifters~\cite{explanation}. 
The interferometer has nine output ports, each of which is connected to a threshold detector $D_{\alpha}$
($\alpha = 1,\dots,9$). 
In other words, the detector $D_{\alpha}$ is clicked if photons live in the $\alpha$th output mode after the transformation (\ref{qutrit_1}).
We regard a triplet ($D_\beta$, $D_{\beta +1}$, $D_{\beta+2}$) as a single detector $\tilde D_{\beta}$ ($\beta = 1,4,7$).
Thus, when at least one element of the triplet is clicked, the logical value of this {\it coarse-graining} measurement is true.
We can find that three distinct events occur at the outputs in the total apparatus:
``bunching" (one of the three {\it coarse-graining} detectors is clicked.),
``anti-bunching" (all the {\it coarse-graining} detectors are clicked.),
and others.
The elements of the corresponding positive operator-valued measure (not
PVM) are 
\(
Q_{1} = (2/3)S_{3}
\) for bunching, 
\(
Q_{2} = (S_{3}/3) + A_{3}
\) for anti-bunching, 
and 
\(
Q_{3} = \I^{\otimes 3} - Q_{1} - Q_{2}
\) for residues,  
with the three-body symmetrizer $S_{3}$. 
Now, we take $\Xi$ as the three-body symmetrizing superoperator, 
\(
\Xi(A) = (1/3!) \sum_{\sigma \in {\mathfrak S}_3}U_{\sigma} A
U_{\sigma}^{\dagger} 
\). 
Then, we find that 
\(
\Xi^{\ast}(M)(x) = \sum_{k=1}^{3} m^{\prime}_{k}(x) Q_{k}
\). 
The polynomials $m^{\prime}_{k}(x)$ can be obtained by straightforward
calculations, and does not depend on $\rho$, like $m_{i}(x)$. 
In this way, we can reconstruct the characteristic polynomial of a qutrit
density matrix, using $m_{k}^{\prime}(x)$ and the measurement probabilities
$\tr (Q_{k} \rho^{\otimes 3})$. 

Finally, we compare our proposal to an approach proposed by Keyl
and Werner~\cite{ref:young}. 
They found a single observable for an eigenvalue determination, via a
group-theoretic approach (see also Ref.~\cite{Hayashi:Matsumoto:2002}). 
Different from ours, the outcome of the observable in a single-shot
measurement is an estimator of the eigenvalues.  
Their method can be considered to be minimal and
implementable in a single setup.
However, to obtain high accuracy, their approach requires an
$N$-i.i.d.\ state $\rho^{\otimes N}$, with $N \to \infty$. 
One has to perform a measurement of a multi-partite
observable in a many-body system, whose particle number depends
on
a given accuracy.
In contrast, 
our observable
is fixed, once
the dimension of a target system is set. 
Thus, our method could be much simpler, from a technical point of view. 

In summary,
we showed a simple method for
measuring the eigenvalues of a density matrix of a $d$-level system in a single setup. 
We also implemented our proposal in linear optical and
superconducting systems.  
In the present formulation, a characteristic polynomial is reconstructed
via quantum measurements.   
The resultant polynomial is straightforwardly calculated by classical
computers. 
This approach is also applicable to evaluating the energy spectrum of a
physical system, like Ref.~\cite{Wang;Nori:2012}. 
Thus, our proposal can be used for a practical assessment of quantum features in a physical system. 

The authors would like to thank A. Miranowicz, S. Ashabb, and S. Tanaka
for helpful discussion. 
T.T. is partially supported by a Grant for Excellent Graduate Schools, MEXT, Japan. 
Y.O. is supported in part by the Special Postdoctoral Researchers Program, RIKEN. 
G.K. and H.N. are partially supported respectively by a Grant-in-Aid for Young Scientists (B) (No. 22740079) and by a Grant-in-Aid for
Scientific Research (C) (No. 22540292) from JSPS.
F.N. acknowledges partial support from the ARO, RIKEN iTHES project, 
JSPS-RFBR Contract No. 12-02-92100, Grant-in-Aid for Scientific Research
(S), MEXT Kakenhi on Quantum Cybernetics, and Funding Program for
Innovative R\&D on S\&T.  


\begin{thebibliography}{99}
\bibitem{footnote1}
One can make an indirect measurement (through the
	Steinspring representation) by preparing an ancilla 
	interacting with the system, and finally performing a projective
	measurement of an observable on the ancilla.
\end{thebibliography}
\end{document}